\documentclass{vgtc}                          

\ifpdf
  \pdfoutput=1\relax                   
  \usepackage{graphicx}                
  \DeclareGraphicsExtensions{.pdf,.png,.jpg,.jpeg} 
\else
  \usepackage{graphicx}                
  \DeclareGraphicsExtensions{.eps}     
\fi%

\graphicspath{{figures/}{pictures/}{images/}{./}} 

\usepackage{microtype}                 
\PassOptionsToPackage{warn}{textcomp}  
\usepackage{textcomp}                  
\usepackage{mathptmx}                  
\usepackage{times}                     
\usepackage{cite}                      
\usepackage{tabu}                      
\usepackage{booktabs}                  

\usepackage{amsmath}
\usepackage{xcolor}
\usepackage{gensymb}

\onlineid{0}

\vgtccategory{Research}

\title{Real-Time Auralization for First-Person Vocal Interaction in Immersive Virtual Environments\thanks{This paper has been accepted for presentation at the 158th Audio\\Engineering Society (AES) Convention, Warsaw, May 2025.}}

\author{Mauricio Flores-Vargas\thanks{e-mail: floresvm@tcd.ie}\\ %
        \scriptsize Trinity College Dublin %
\and Enda Bates\thanks{e-mail: ebates@tcd.ie}\\ %
     \scriptsize Trinity College Dublin %
\and Rachel McDonnell\thanks{e-mail: ramcdonn@tcd.ie}\\ %
     \parbox{1.4in}{\scriptsize \centering Trinity College Dublin}}

\abstract{Multimodal research and applications are becoming more commonplace as Virtual Reality (VR) technology integrates different sensory feedback, enabling the recreation of real spaces in an audio-visual context. Within VR experiences, numerous applications rely on the user’s voice as a key element of interaction, including music performances and public speaking applications. Self-perception of our voice plays a crucial role in vocal production. When singing or speaking, our voice interacts with the acoustic properties of the environment, shaping the adjustment of vocal parameters in response to the perceived characteristics of the space. 
 
This technical report presents a real-time auralization pipeline that leverages three-dimensional Spatial Impulse Responses (SIRs) for multimodal research applications in VR requiring first-person vocal interaction. It describes the impulse response creation and rendering workflow, the audio-visual integration, and addresses latency and computational considerations. The system enables users to explore acoustic spaces from various positions and orientations within a predefined area, supporting three and five Degrees of Freedom (3Dof and 5DoF) in audio-visual multimodal perception for both research and creative applications in VR.
} 

\begin{document}
\firstsection{Introduction}
\maketitle
Multimodal research in Virtual Reality (VR) has gained significant attention due to the possibility of VR technology to provide both audio and visual feedback \cite{Malpica2020}. Studies have explored aspects such as immersion, presence, effects on users' physical and mental states, and audio-visual perception in immersive environments \cite{Martin2022, Kern2020, Buetler2022, Mahmud2022, Tajadura2018, Senna2014, Tajadura2015}. Moreover, it has enabled the development of multi-sensory experiences that take advantage of this integration \cite{Geronazzo:2022:SIVE}, including multimodal reconstruction of acoustic environments, music performances and public speaking \cite{Kearney2016, Luizard2019, Poeschl2017}.

Self-perception of one's voice significantly influences vocal production. Whether in singing or speaking, the acoustic characteristics of a space (e.g. early reflections and diffuse reverberation) play a major role in vocal perception, adaptation, and performance, particularly affecting timbre and sound level \cite{Luizard2019, Redman2023}. Acoustic simulation using three-dimensional Spatial Impulse Responses (SIR) captured using binaural and Ambisonic techniques is a popular method for sound auralization \cite{Kearney2016, Luizard2019}. Previous work has focused on the recreation of real-world spaces to create virtual immersive experiences for listeners. However, they are mostly captured from a third-person perspective, meaning that the listener is not the performer (i.e. the source), making them unsuitable for self-perception of vocal interaction.

This work is a case study report on the design and implementation of a pipeline that enables first-person real-time sound auralization in VR. The system allows users to perceive the acoustic response at different positions and directions within a defined area, providing audio-visual feedback for multimodal studies and practical applications.
The pipeline has been split into two systems: the production of SIRs from a first-person perspective and the audio signal processing to auralize the user's voice in real time within Immersive Virtual Environments (IVE).

\section{SIR Production}
The SIR production closely follows the methods proposed by Kearney et al. \cite{Kearney2022} and Mroz et al. \cite{Mroz2021}. While the dataset from Kearney et al. \cite{Kearney2022} includes four SIRs at the performer’s position, they were recorded from non-adjacent locations, limiting their application to three degrees of freedom (3DoF). Therefore, the proposed system is an adaptation designed to capture various adjacent positions and directions, providing the flexibility for users to teleport or move freely within an area, thus enabling five degrees of freedom (5DoF): three rotational (roll, pitch, and yaw) and two translational (left-right and forward-back) movements.  

The system is limited to two translational degrees of freedom, as the height remains fixed across all recordings. Capturing the third translational DoF would require additional layers of impulse responses recorded at varying heights to account for elevation \cite{Patricio2019}, which would incur significant costs in terms of capture and reproduction. Therefore, we deemed it unnecessary, as performers typically maintain a relatively stable vertical position, and this approach has been shown to be effective in similar work \cite{Kearney2016, Mroz2021}.

\begin{figure}[ht]
\begin{center}
\includegraphics[width=0.98\columnwidth]{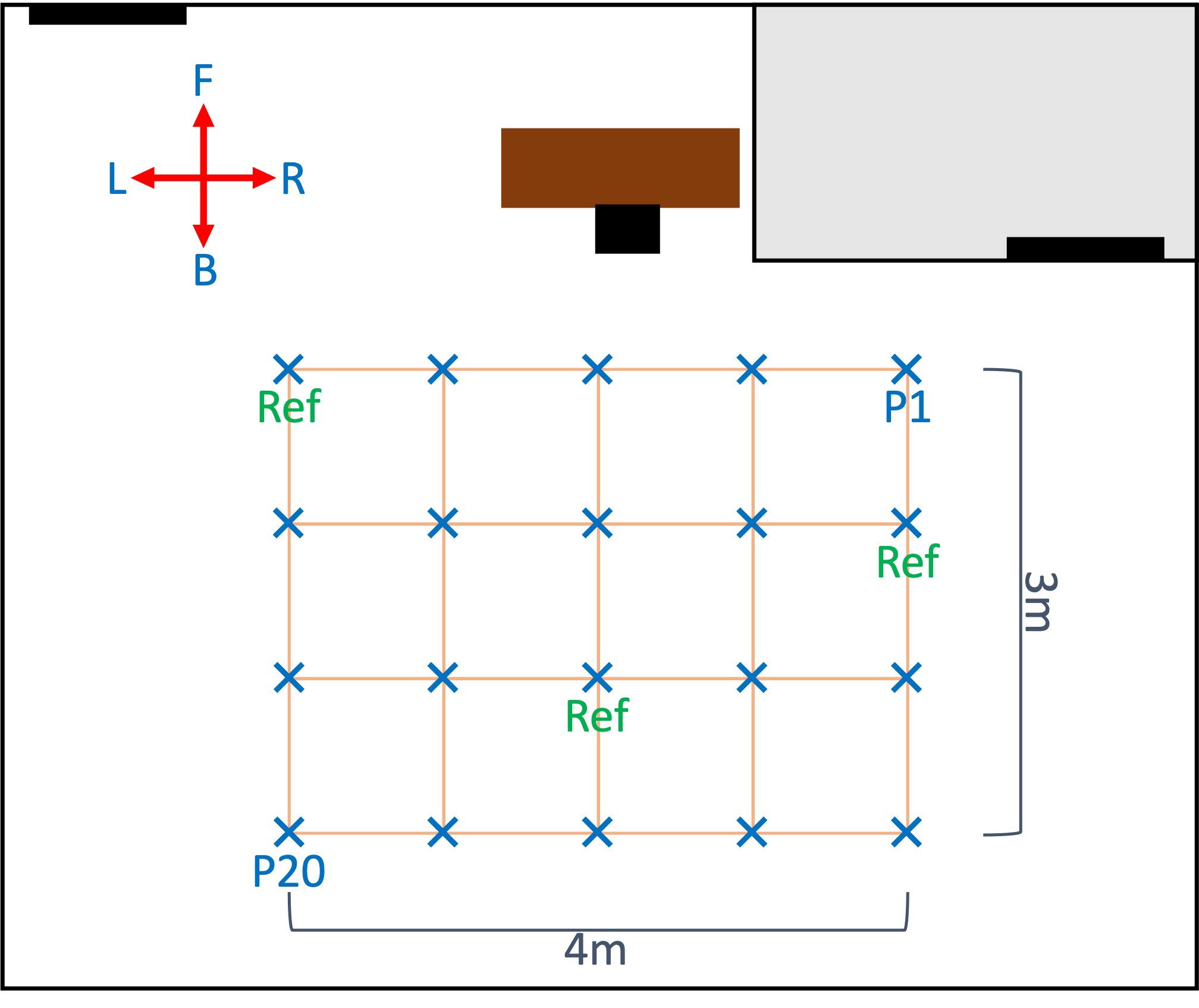}
\caption{Grid of recording positions. }
\label{fig:positions}
\end{center}
\end{figure}

\paragraph{\textbf{Pre-recording Setup}}
To allow for translational DoF, a grid comprised of 20 recording positions spaced 1 meter apart were defined, providing a total area of 12m\textsuperscript{2} (4 x 3 meters) for user exploration (See Figure \ref{fig:positions}). Initially, a triangular distribution, as used by \cite{Mroz2021}, was considered since it is less computationally intensive; however, this arrangement results in a smaller area for the same number of positions (approx. 9m\textsuperscript{2} for 20 positions spaced 1 meter apart) with an irregular shape.

\begin{figure}[ht]
\begin{center}
\includegraphics[width=0.36\columnwidth]{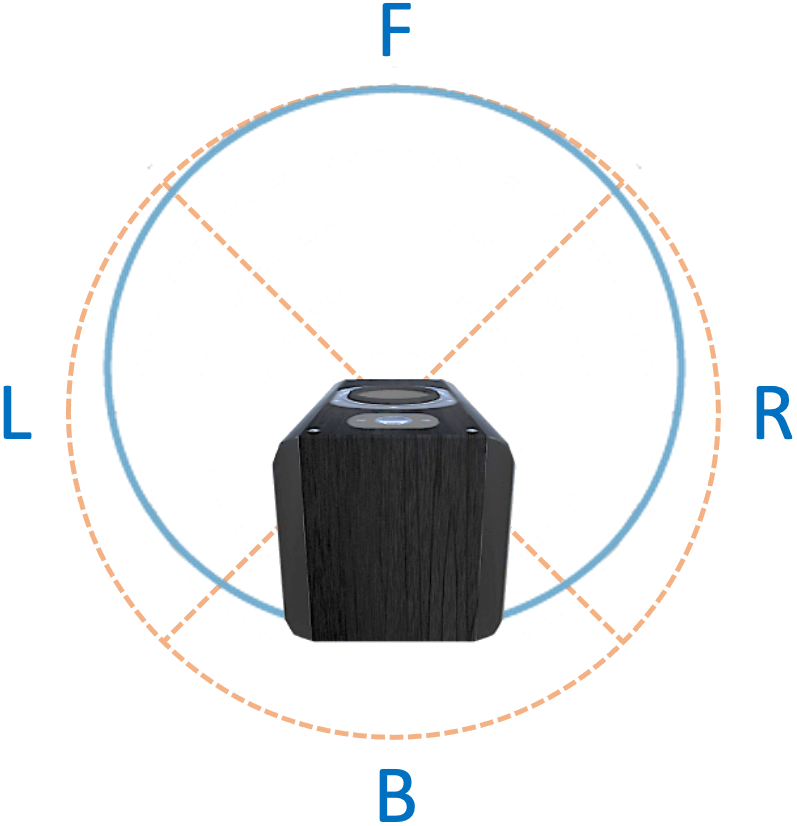}
\caption{Speaker directivity pattern.}
\label{fig:rotations}
\end{center}
\end{figure}

To enable rotational DoF, four directions were considered for each position: front, right, back, and left. This approach was necessary because the speaker used (Genelec 1029A) to excite the environment has a somewhat subcardioid pattern \cite{Canclini2015}, comparable to the directivity of the human voice. As a result, the acoustic response varies depending on the speaker's orientation within the acoustic space (See Figure \ref{fig:rotations}).

A Digital Audio Workstation (DAW) project was set up to streamline the recording process. The project included a mono audio track set at -20dBFS containing 80 sine sweeps, each 20 seconds long, corresponding to 20 positions with 4 directions each. Preceding each sine sweep, the project included a voice identifier that specified the respective position and direction followed by a sync-pop to ensure synchronization during subsequent processing, and four reference recordings were included for reference against the convolved signal outputs. A 5-second interval followed each sine sweep to allow the reverberation to decay before the next recording. In addition, the project included a 20-channel track to record the input signal from the 19 omnidirectional condenser microphone capsules of the Zylia ZM-1 microphone \cite{ZyliaMAN}. These recordings were subsequently encoded into 3rd-order Ambisonics (3OA) using the Zylia Ambisonics Converter plugin.

\paragraph{\textbf{Recording Process}}
Acoustic recordings were performed with a third-order Ambisonic (3OA) microphone (Zylia ZM1) to capture acoustic responses within a 360$^{\circ}$ sound field. The speaker was set at a fixed height of 1.5m, with the microphone placed 2cm above, simulating the relative position of the mouth and ears of a person and representing a first-person perspective. The speaker output was calibrated to 85dBA 
using an SPL meter placed 1 meter away. The microphone gain was then adjusted to peak around –3dBFS to prevent clipping caused by resonances in the acoustic space. During the recording process, both the speaker and microphone were placed at each designated position. The microphone was consistently oriented to the front of the stage, while the speaker was rotated sequentially to face each of the four directions: front, right, back, and left.

\paragraph{\textbf{SIRs Creation}}
After completing the measurement recordings, a single 16-channel WAV file (3OA) was exported. A MATLAB script was developed to segment the file into 80 individual measurements, using the sync-pops for precise segmentation. Each channel was exported separately, resulting in 16 mono audio files per measurement. These mono files were deconvolved using Voxengo's Deconvolver \cite{VoxengoDec}, chosen for its reversed test tone deconvolution technique, which effectively mitigates artifacts and improves the quality of impulse responses. Processing was conducted on mono files as the software is limited to a maximum of 8 channels. 
Finally, all deconvolved files were recombined into 16-channel SIRs.

\section{Auralization Pipeline}
\begin{figure}[ht]
\begin{center}
\includegraphics[width=0.98\columnwidth]{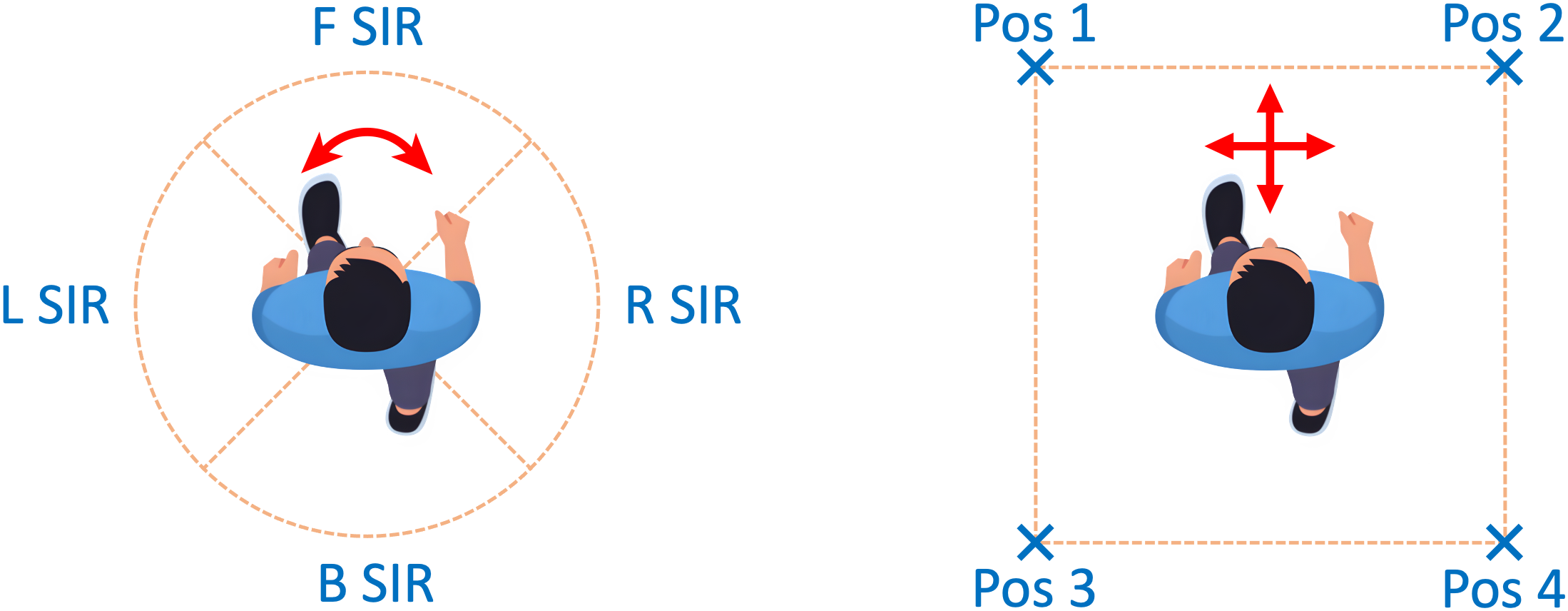}
\caption{Directional (left) and Translation (right) interpolation.}
\label{fig:interpolations}
\end{center}
\end{figure}

This pipeline is responsible for auralizing the input signal, specifically the performer's voice, based on the user’s position and rotation within the virtual environment. To achieve the necessary level of freedom, the system interpolated between 16 different SIRs, which are convolved in real-time with the input signal from the microphone. Therefore, the interpolation happens between 4 positions with 4 different directions (See Figure \ref{fig:interpolations}).

\subsection{System Setup}

\paragraph{\textbf{Direct Signal Amplitude Compensation}}
Noise-canceling headphones were used to block the existing acoustic response of the physical environment. While some studies propose using open-back headphones to allow direct sound to reach the musician \cite{Accolti:2023:RTA}, we opted for noise-canceling headphones based on the fact that, unless users were in an acoustically transparent space such as anechoic chamber or heavily treated room, open-back headphones would result in hearing a mix of both the virtual auralization and the physical room’s acoustic response. However, using headphones, especially noise-canceling ones, results in a perceived amplitude loss of the user’s voice. To account for this, the direct input of the microphone was fed directly to the headphones via the audio interface’s direct monitoring feature, which allows listening to the input signal with near-zero latency. The amplitude loss compensation was calculated by comparing the signal drop of white noise recordings conducted using a Sennheiser KU100 dummy head microphone, with and without headphones placed on the dummy head’s ears, and a speaker positioned 1 meter away facing the microphone.

\paragraph{\textbf{Latency Compensation}}
One of the biggest challenges in self-produced sound processing is the inherent latency between producing a sound and hearing its processed output. This latency is not fixed and arises from multiple factors, including the audio interface and the system digital processing and buffering before it reaches the headphone output. Therefore, we aimed for the lowest buffer size possible without performance glitches or dropouts.

To overcome this issue, we cropped the direct sound and the Initial Time Delay Gap (ITDG) of the SIRs up to the first reflection and then aligned them using a delay plugin, with the delay time compensation obtained by subtracting the round-trip latency from the ITDG. Although significant early reflections were apparent in the impulse responses, the ITDG to the nearest surface (the ceiling, in this case) was estimated using using the following expression:
\[
\Delta t = t_1 - t_0 = \frac{(L_1 + L_2) - L}{c}
\]
where \(t_0\) and \(t_1\) are the arrival times of the direct and reflected paths respectively, \(L\) is the direct path length, \(L_1 + L_2\) is the reflected path length, and \(c \approx 344\,\mathrm{m/s}\) is the speed of sound \cite{EpifaniITDG}.

Most commonly, the first arriving reflection comes from the floor as it is the closest surface to the performer. However, similar to \cite{Kearney2016}, we did not consider this reflection due to its negligible significance, attributed to the shadowing effect of the singer's torso and the voice's directivity. Nonetheless, we acknowledge that different scenarios may require estimating the ITDG based on the floor reflection.

\begin{figure}[ht]
\begin{center}
\includegraphics[width=0.98\columnwidth]{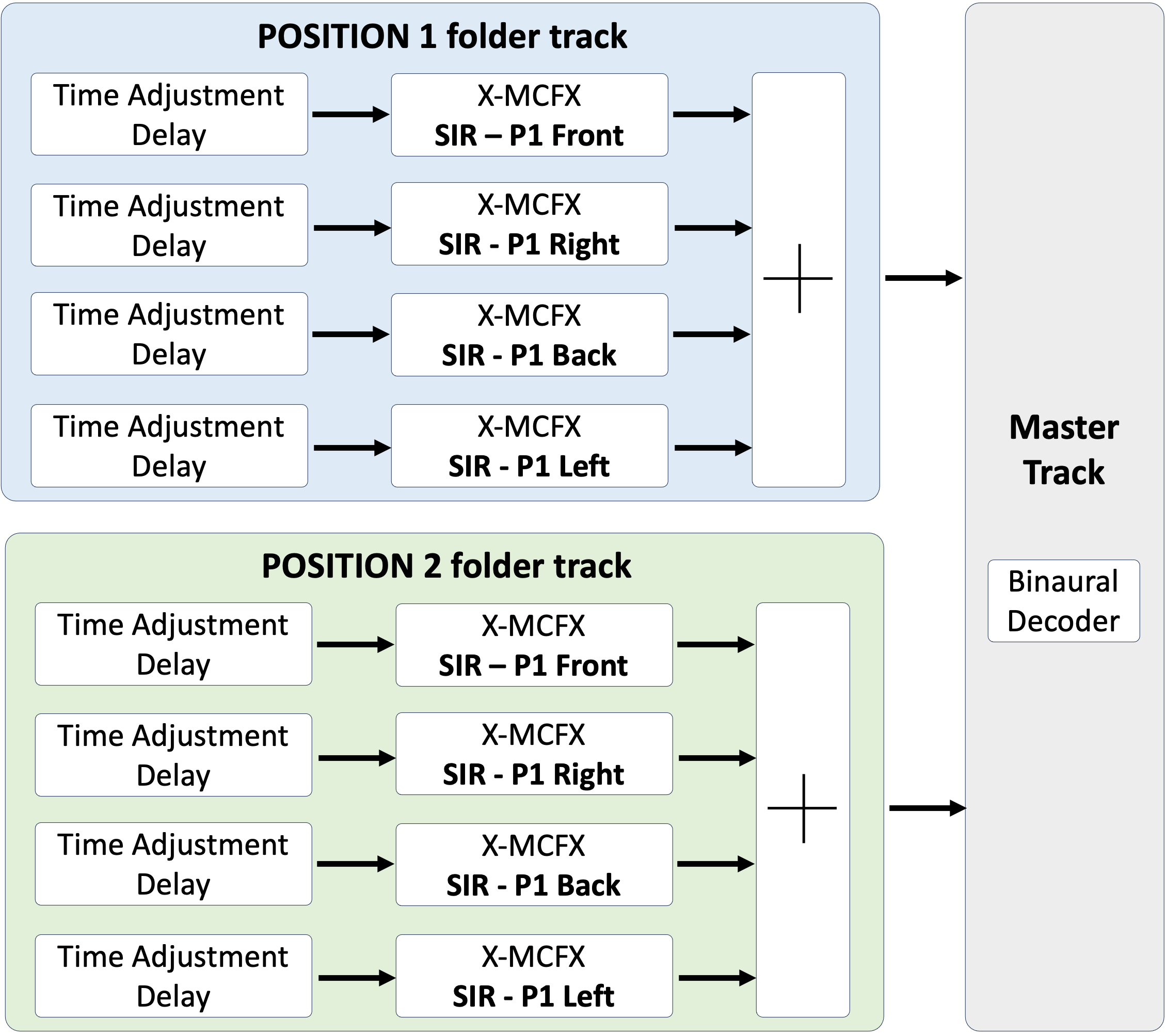}
\caption{Signal flow showing two of the twenty folder tracks in the Reaper project.}
\label{fig:reapersignalflow}
\end{center}
\end{figure}

\paragraph{\textbf{DAW Project Setup}}
The audio signal processing was handled within Reaper DAW (See Figure \ref{fig:reapersignalflow}). The project was organized using folder tracks. Each folder track corresponded to one position on the area grid and contained four 16-channel audio tracks representing each direction (front, right, back, left). Each direction tracks included a time adjustment delay plugin 
and a multichannel convolution plugin loaded with the corresponding SIRs as inserts. 
The microphone input was assigned to a single 16-channel track containing an Ambisonic encoder plugin to encode the mono signal into 3OA, which was then routed to the directional tracks of all positions for convolution. Finally, all position folder tracks were routed to the Master track, which contained a binaural decoder plugin to decode the Ambisonic signal into a binaural headphone signal.

\subsection{Real-time Signal Interpolation}
The signal auralization was achieved by interpolating the convolved microphone input with the SIRs across two independent gain stages: Directional and Translation interpolation.

\paragraph{\textbf{Directional Interpolation}}
Interpolation between the four directional SIRs (front - right -back - left) of each position was performed using constant power panning to adjust the gain of each corresponding audio track in the Reaper DAW, based on the HMD's orientation relative to the IVE.

Since the VR experience was rendered in Unity, a C\# script was developed to manage the interpolation. The script continuously tracked the HMD’s rotation each frame, wrapped the values to a 0–360$^{\circ}$ range to avoid negative angles, and applied 4-way constant-power panning based on the yaw angle. Afterwards, the resulting four gain values were transmitted to Reaper via the OSC protocol, controlling the directional tracks. The gain $G_d$ for each direction $d$ was calculated using cosine-based constant power panning. Since power is proportional to the squared amplitude, and $\cos^2(\theta) + \sin^2(\theta) = 1$, assigning gains as $L(\theta) = \cos(\theta)$ and $R(\theta) = \sin(\theta)$ provides constant power panning \cite{OlandLCPA}.

\paragraph{\textbf{Translational Interpolation}}
Previous work has explored different interpolation strategies for 6DoF audio rendering \cite{Patricio2019, Alary2021, McCormack2022}. In the transitional interpolation stage of this pipeline, we opted to implement Inverse Distance Weighting (IDW) due to its flexibility with node distribution, perceptual robustness, and computational efficiency.
IDW computes a weighted average of surrounding node values, where each weight is inversely proportional to the node’s distance from the listener \cite{Alary2021}. The gain interpolation was defined as:
\[
w_i = \frac{\frac{1}{d_i}}{\sum\limits_{i=1}^{M} \frac{1}{d_i}}
\]
where $d_i$ is the distance from the performer to the $i$th impulse response position, and $M$ is the number of contributing nodes.

A second C\# script was developed to perform IDW-based interpolation between positions, using the HMD’s location within the immersive virtual environment (IVE). The script dynamically tracked the user’s position and computed normalized weights based on the inverse distance to active IR nodes. These values were then transmitted to Reaper via OSC to control the amplitude of the corresponding folder tracks. To improve functionality and performance, the script also limited the number of active nodes and disabled audio plugins in the Reaper project when a node was inactive, using hysteresis thresholds to prevent unintended on/off toggling.

\section{Conclusion}
In this paper, we presented the design and implementation of a first-person vocal auralization system that enables real-time 5DoF vocal interaction within immersive IVEs. The pipeline uses 3OA SIRs captured from a first-person perspective to deliver 360$^{\circ}$ acoustic representation of the space. We describe in detail the process of impulse response generation as well as the real-time amplitude interpolation across both rotational and translational dimensions. This system allows users to explore acoustic spaces in 5DoF, successfully supporting audio-visual perception in both multimodal research and creative applications in IVEs.


\end{document}